\documentstyle[manuscript,aps]{revtex}

\newcommand{\beq}{\begin{equation}}
\newcommand{\eeq}{\end{equation}}
\newcommand{\beqa}{\begin{eqnarray}}
\newcommand{\eeqa}{\end{eqnarray}}
\newcommand{\beqar}{\begin{eqnarray*}}
\newcommand{\eeqar}{\end{eqnarray*}}

\def \A {{\bf A}}
\def \p {{\bf p}}
\def \x {{\bf x}}
\def \Q {{\bf Q}}
\def \P {{\bf P}}

\def \xzero {{\bf x}^{(0)}}

\def \xone  {{\bf x}^{(1)}}

\def \xtwo  {{\bf x}^{(2)}}

\def \hzero {\Delta H^{(0)}}
\def \hone  {\Delta H^{(1)}}
\def \htwo  {\Delta H^{(2)}}

\def \xf    {{\bf x}_0}
\def \pf    {{\bf p}_0}

\def \half {{1\over2}}

\begin{document}

\title{ \bf\Large  
Measurability in Linear and Non-Linear \\
 Quantum Mechanical Systems}

\author{
{\large Y. Aharonov$^{(a)}$  and B. Reznik$^{(b)}$
\footnote{\it e-mail: reznik@t6-serv.lanl.gov\\
LAUR-97-186}  }\\
{\ }\\
      $^{(a)}$ {\it School of Physics and Astronomy,
 Tel-Aviv University, Tel-Aviv 69978, Israel}\\              
    {\it and Department of Physics,
     University of South Carolina, Columbia, SC  29208} \\
{\ } \\
       $^{(b)}$ {\it Theoretical Division T-6 MS B288,  
Los Alamos National Lab., Los Alamos, NM 87545} 
} 

\maketitle 

\begin{abstract}
The measurability by means of  continuous 
measurements, of an observable $\A(t_0)$, at an instant,
and of a time averaged observable, $\bar \A=1/T\int \A(t')dt'$,  
is examined for linear and in particular for 
non-linear quantum mechanical systems.
We argue that only when the exact (non-perturbative) solution
is known, an exact measurement may be possible. 
A perturbative approach is shown to 
fail in the non-linear case for measurements with accuracy $\Delta \bar
\A < \Delta \bar \A_{min}(T)$, giving rise to a restriction
on the accuracy.
Thus, in order to prepare an initial pure state of a 
non-linear system, by means of a continuous measurement, 
the exact non-perturbative solution must be known. 
\end{abstract}
\newpage


\section{Introduction}

The measurement of an  
observable $\A(t_0)$ is ideally described by an 
impulsive coupling between the system and the measuring
device at $t=t_0$ \cite{vn}. 
In a realistic measurement however, due to the finite duration 
of the coupling between the systems, the back-reaction of the measuring 
device modifies the free evolution of the system. 
Thus, one generally observes a time averaged observable
which depends of {\it both} the system and the measuring device. 
This raises the following questions:
\begin{itemize}
\item  Can the  {\it exact} value
of $\A(t_0)$ at the instant $t=t_0$, 
be measured by means of a {\it continuous} measurement?
\item  Can the value of a time averaged 
observable, $\bar \A = T^{-1}\int_0^T \A(t')dt'$, be 
precisely measured? 
\end{itemize}
In this article we shall address these questions in the context
of linear and in particular of non-linear quantum mechanical systems.
Other aspects of continuous measurements have been discussed
for example in Refs. \cite{mensky,peres,caves,hartle}.

The answer to the first question above has fundamental significance  
for quantum mechanics. According to quantum mechanics, in order to 
{\it prepare} a certain initial pure state $\psi(t=t_0)$, one has to 
determine the values of a complete set of observables
at the {\it instant} $t=t_0$. But what if only continuous measurements
are  possible?

It turns out that for linear as well as non-linear systems, 
as long as the $exact$ solution to the Heisenberg equations
of motion of the system is known, such a preparation 
is at least formally possible. 
To see this, 
consider a general exact solution for the observable $\A(t)$: 
\beq
\A(t) =  \xi(t-t_0,\A_m(t_0)),
\eeq
which is given in terms of initial values 
of the observables $\A_m$
at time $t=t_0$.
By sending $t\to -t$ and replacing the roles of $t$ and $t_0$
we can represent the constant of motion, $\A(t=t_0)$, 
in terms of time dependent operators $\A_m(t)$:
\beq
\A(t=t_0) = \xi(t_0-t, \A_m(t)).
\eeq
We can now use the standard continuous measurement described by
the interaction 
\beq
H_I = -g(t) \Q \xi(t_0-t, \A_m(t)),
\eeq
where $\Q$ is conjugate to the ``output observable'' $\P$,
and $g(t)=g_0$ for $t_0<t<t_0+T$ (or more generally, $g$ has   
a finite support in time, and $\int g dt$ is finite). 
Since $[\A(t_0), H_I]=0$, the measurement does not 
cause an error in $\A(t_0)$, the exact 
value can be measured.
However, this approach requires the knowledge of the exact
non-perturbative solution.  But what if the latter is 
unknown? We shall further discuss this issue  in Sections 5-7.

The second question raised above becomes particularly acute 
when asked in the context of 
quantum field theory.  
The singular nature of a field at a point requires 
to consider space-time weighted (smeared) fields
as the elementary observables of quantum field theory
\cite{haag}. Therefore, we are faced with the problem of 
measuring a space-time averaged observable.
The measurability of such objects has been investigated 
long ago by Bohr and Rosenfeld\cite{br} for the
case of a free field, but has not been considered
for non-linear field theories \cite{aharonov-reznik}. 

As a toy model we shall consider in this article 
the problem in the simpler case of non-relativistic quantum theory.
For the case of a linear theory, 
the back-reaction on the system during a measurement of 
$\bar \A=T^{-1}\int \A(t')dt'$
depends only on variables of the measuring device. 
A compensation term can therefore be devised so as to 
exactly compensate for the back-reaction, and measure the
latter averaged observable to any desired accuracy.
But if the system has non-linear equations of motion, 
the problem of observing a time averaged quantity 
becomes much more difficult. 
The error due to the non-linear back-reaction 
becomes a function of both system and measuring device variables.
As already described above, if a non-perturbative solution is known,  
then formally a precise measurement is possible. 
Since in general however this is not the case, 
we shall consider in some details the problem of 
measuring a time averaged observable by means of  
a perturbative approach. We shall show that 
the validity of a perturbative method (in the non-linear coupling
constant) is limited. For measurement with of accuracy 
better than a certain minimal uncertainty, the perturbation 
scheme breaks-down. 
 
In the next section we shall
present the toy model which will be used in this article
to study linear and non-linear cases: a harmonic
oscillator with a non-linear potential. 
In Section 3. we shall consider three different 
approaches to measure averaged observables in the linear case.
We then discuss these methods when the system is non-linear 
in Section 4.  A perturbative approach to the non-linear case 
is developed in Section 5, and shown to break-down for precise
measurements in Section 6. We conclude 
by a discussion of the main results.

\section {Non-linear harmonic oscillator as a toy problem}

We shall consider as a toy model for a non-linear 
system a harmonic oscillator with
a potential given by
\beq
H = \half (\p^2 + \Omega^2 \x^2)  - {\lambda\over n } \x^n, 
\label{model}
\eeq
where $n>2$.

Our first aim will be to describe a measurement of an averaged
observable such as
\beq
\bar \x  = {1\over T} \int_0^T \x(t') dt'.
\eeq
In the limit of $T\to 0$ the prescription is know:
the appropriate interaction term is in this case 
\beq
H_I = - g(t) \Q \x, 
\label{vn}
\eeq
where $\Q$ is canonically conjugate to the "output
variable" $\P$,
and $g(t)= g_0\delta(t)$. It is assumed that the effective 
mass of the device is infinitely large, thus the kinetic part 
of the measuring device Hamiltonian vanishes and  
$\Q$ is a constant of motion. 
The interaction Hamiltonian (\ref{vn}) yields:
\beq
\delta{\bf \P} \equiv \P(+\epsilon) - \P(-\epsilon) = g_0 \x(0). 
\eeq

As a starting point let us modify this interaction by replacing 
the Dirac-delta function by a smooth function $g(t)$ with 
a compact support only the time interval $0<t<T$.

The solution to the equations of motion
$$
\dot \x = \p
$$
$$
\dot \p = -\Omega^2 \x + \lambda \x^{n-1} + g(t)\Q 
$$
\beq
\dot\P = g(t) \x, 
\eeq
can be written  in the integral form:
\beq
\x(t,g) = \x_0(t)  + 
\int_0^t {F(t')\over \Omega} \sin[\Omega(t-t')] dt'
\label{intrep}
\eeq
\beq
\p(t,g) = \p_0(t) + \int_0^t F(t') \cos[\Omega(t-t')] dt' , 
\eeq
where
\beq
F(t) = {\lambda} \x^{n-1}(t,g) + g(t) \Q,
\eeq
and  $\x_0(t), \ \ \p_0(t)$ are free solutions ($\lambda=g=0$),
which coincide with the non-linear solution $\x(t)$
at $t=0$. The presence of the coupling 
parameter $g$ in $\x(t,g)$ will be used in the following to 
denote an explicit
dependence of the solution on variables of the measuring device.

The solutions  $\x(t,g)$ and $\p(t,g)$ depend on the 
back-reaction via the term $g(t) \Q$. 
Contrary to classical mechanics, 
due to the uncertainty principle, we can not make 
$\Q$ as small as we wish and still obtain an accurate
measurement, 
i.e. $\Delta \P(0) \to 0$.
Consequently,  we finally observe  
\beq
\P(T)- \P(0) = \int_0^T g(t)\x(t,g)dt , 
\eeq
which is not the  undisturbed (g=0) value of $\bar \x$.
Only in the limiting case $T=0$ does the error vanish.

In next section we show how 
the undisturbed $\bar \x$ can be observed  
in the linear case; $\lambda=0$.

\section{ The linear case}

We shall now proceed to examine the case of a measurement of 
an averaged observable and show how to 
eliminate the back reaction in this  special case.

For simplicity we shall choose the weight function as 
$g(t)= g_0$ for $t\in (0,T)$ and 
zero otherwise. In this case, the 
solution of the equations of motions when $\lambda=0$ is
$$
\x(t) = \x_0(t) + {g_0(1-\cos\Omega t) \over \Omega^2} \Q
$$
\beq
\equiv \x_0(t) + \alpha(t)\Q . 
\label{alpha}
\eeq
Thus, the $\P$ coordinate of the measuring device will 
be shifted by
\beq
\delta\P = \P(T) - \P(0) = g_0\int^T_0 \x(t')dt'= 
g_0 \int_0^T \x_0(t')dt'  +  \Q {g_0^2 T\over \Omega^2}
\biggl( 1 - {\sin \Omega T\over \Omega T} \biggr) .
\label{deltap}
\eeq
The undisturbed average $\bar \x =\bar \x_0 =
T^{-1}\int_0^T \x_0(t')dt'$  is given by
\beq
\bar \x_0 = {\delta\P\over g_0T} -  
  {g_0 \over \Omega^2}
\biggl( 1 - {\sin \Omega T\over \Omega T} \biggr) \Q . 
\label{x0}
\eeq

Since $\bar \x_0$ depends linearly on both
$\P$ and $\Q$, a precise measurement of $\P$ causes a 
larger uncertainty 
in the second term above. I.e. it increases the back reaction of
the measuring
device on the oscillator. 
Since the uncertainty in $\bar \x$ is
\beq
\Delta \bar \x_0 \approx  {\Delta \P\over g_0T} +  
 {g_0 \over \omega^2 \Delta\P}
\biggl( 1 - {\sin \Omega T\over \Omega T} \biggr) ,
\eeq
the minimal uncertainty is 
\beq
\Delta\bar x_{min} = {2\over \Omega}\sqrt{1-{\sin\Omega
T\over\Omega T}}
\stackrel{{T\to 0}}{\to}
\sqrt{2\over3} T , 
 \eeq
which vanishes only in the impulsive limit. We also note that
in the limit
of $T\to \infty$ the disturbance does not average out but
rather approaches the constant ${2\over \Omega}$.

The direct approach therefore fails to measure $\bar \x_0$
precisely.
There are however ways to correct or eliminate the error above,
and in the following we present three different ways
to achieve this goal. 

The idea of Bohr and Rosenfeld was to  
correct the error by adding to the 
Hamiltonian (\ref{vn}) a new ``compensating'' term.
The error in the shift of $\P$
in eq. (\ref{deltap}) appears as linear in $\Q$, very much like 
the effect of a spring in the $\Q$-coordinates.
Since the coefficient which multiplies $\Q$ in (\ref{deltap})
is known, it is straightforward to compensate for this error
by adding to the interaction Hamiltonian (\ref{vn})
a spring term:
\beq
H_I = - g(t) \Q \x + \half k \Q^2 , 
\eeq
where $k$ can be chosen for example as  
\beq
k(t) = {g^2(t)T\over \Omega^2 }
\biggl( 1 - {\sin \Omega T\over \Omega T} \biggr) .
\label{k}
\eeq
It is straightforward to see that the new spring-term 
in the equations of motion precisely eliminates the 
back-reaction of the device on the system.
In the limit $T\to 0$, the compensation is of course not 
needed since the spring constant vanishes, and we obtain
back the ordinary impulsive measurement.

Another approach due to 
Unruh \cite{unruh}, does not require 
modification of the interaction (\ref{vn}) but requires 
instead an additional measuring device. 
Inspecting eq. (\ref{x0}) we notice that $\bar \x_0$ is given
on the right hand side in terms of a linear combination
of $\P$ and $\Q$. Therefore, 
after preparing the measuring device in an initial state with 
a well defined $\P$ one can measure
by means of {\it another} measuring device
the linear combination 
\beq
 {1\over g_0T}\P  - {k(T)\over g_0} \Q(T)  ,
\eeq
where $k$ is given in eq. (\ref{k}).
As before this method requires that the back-reaction 
effect on the system depends only on variables of the 
measuring device.
As we shall see in the next section, 
both methods apparently fail when non-linearities
introduce back-reaction terms which depend also on the system
itself.

A third different method to measure $\bar\x$ 
could be the following.
By integrating $\x(t)$ we can express $\bar\x$ as
\beq
{1\over T}\int_0^T(\x_0\cos\Omega t' + {1\over
\Omega}\p_0\sin\Omega t')dt'
= {\x_0\over \Omega T}\sin\Omega T + {\p_0\over \Omega^2 T}
(1-\cos\Omega T) .
\eeq
Note that  $\x_0$ and $\p_0$ are constants of motion. 
By inverting the solutions for the equations of motion 
for $\x=\x(\x_0,\p_0)$ and  $\p=\p(\x_0,\p_0)$ we 
obtain 
\beq
T \bar\x(t)= {2\over \Omega}\sin\Omega {T/2}\biggl[
 \x(t) \cos\Omega({T/2}-t)  + 
{1\over \Omega} \p(t)\sin\Omega({T/2}-t) \biggr] .
\eeq
Thus,  we have expressed $\bar \x$ in terms of a (time dependent)
constant of motion. 
We can now set up a continuous measurement
\beq
H_I = -g(t) \Q \bar\x(t).
\eeq
Clearly  the observable $\bar\x(t)$
remains a constant of motion  
and is not disturbed by the interaction.
We have thus managed to measure  
$\bar\x$ by a continuous measurement 
of a constant of motion whose value 
is identical to $\bar \x$ for any $t$.

\section{ Non-linear case}

The presence of non-linear  terms makes the back-reaction 
dependent on both system and 
measuring device variables. This makes the problem of
constructing the analogous compensation more complicated.

Nevertheless, if the exact solution is known, there is at least
a formal way of eliminating completely the back-reaction.
To this end we can use the third approach that was described
in the last section.
Let use denote the exact solutions of the 
equation of motion by
\beq
x(t) = \xi(\x_0,\p_0,t), \ \ \ \ p(t) = \xi'(\x_0,\p_0,t),
\eeq
Then, the average $\bar\x$ may be expressed as
\beq
T\bar\x = \int_0^T \xi(\x_0,\p_0,t')dt'  = \phi(\x_0,\p_0,T) ,
\label{nlconstant}
\eeq
where the latter observable, $\phi$, is a constant of motion. 
We can now substitute
\beq
\x_0 = \xi(\p(t),\x(t),-t), \ \ \ \ \p_0= \xi'(\p(t),\x(t),-t), 
\eeq
into  eq. (\ref{nlconstant}) and express $\phi$
in terms of the time dependent dynamical variables:
\beq
\bar \x = \phi\Bigl(\xi(\p(t),\x(t),-t),\xi'(\p(t),\x(t),-t),T\Bigr)
= \phi(\x(t),\p(t),t,T) .
\eeq
Since $\phi(\x(t),\p(t),t,T)$ is a constant of motion we can 
measure it by means of the ordinary interaction (\ref{vn}).

The other two methods of Section 3. are much harder to use. 
Assume again the exact solution is given, and let us express it
as
\beq
\x(t,g) =  \xi(t,\lambda,g=0) +\Delta\xi(t,\lambda,g) . 
\eeq
Here, $\Delta\xi(t,\lambda,g)=\xi(t,\lambda,g)-
\xi(t,\lambda,g=0)= x(t,g) - x(t,g=0) $
is the non-linear 
error due to the measurement. 
Thus, using the naive coupling (\ref{vn}) we obtain:
\beq
{\delta\P\over g_0T} = \bar  \x 
+{1\over g_0T} \int_0^T \Delta\xi(t,\lambda,g)dt' .
\eeq
To proceed, we would like to construct a compensation to 
the last term.\footnote{
At this point we note that 
in principle we could have arranged that 
during the measurement a compensation term 
$-g(t)\lambda \x^n/n$ eliminates the non-linearity.
We could then use the previous method for observing 
the time averaged observable corresponding to a
a linear dynamics.}
Thus, $\Delta\xi$ must be expressed in terms of the
dynamical variables
$\x(t,g)$, $\p(t,g)$ and $\Q$. However, $\Delta\xi$ 
depends on $\x(t,g=0)$ and we first need to find the 
non-linear relation between the disturbed non-linear $\x(t,g)$
and the undisturbed $\x(t,g=0)$ solutions:
\beq
\x(t,g) = f[\x(t,0), \p(t,0),\Q],
\label{relation}
\eeq
etc. The latter equation corresponds to 
a ``canonical'' transformation between two non-linear solutions
of two different systems.  It is not clear if such a well defined
relation indeed exist.  
Assuming however that transformation above does exist, 
we may now attempt, as before,
to add to the Hamiltonian the compensation term
$\int [k\Q + \Delta\xi(\lambda,g)]dq$. However,  
the new equations of motions will give rise to another
non-linear error $\Delta\xi'\ne\Delta\xi$. 
Therefore, a self consistent scheme  
for  constructing a compensation must be found, which takes
deals with the non-perturbative effects
of the compensation on the system. 

The difficulties in finding an exact 
non-perturbative compensation scheme, leads us to examine a more 
modest approach. 
It can be  hoped that at least for small averaging time $T$, 
the effect of the non-linearities is small. Therefore, 
in the next section, we shall examine a perturbative approach.

\section{A perturbative approach}

For a small non-linear constant,  $\lambda$, 
we shall attempt a  perturbative approach.
As a starting point we shall assume that the 
measurement can be expressed by:
\beq
H_I = -g(t)\Q\x + H_C , 
\eeq
i.e., as a sum of the naive measurement and a compensation term.
Clearly in the limit of $T\to0$, $H_I$ should reduce to an 
ordinary impulsive 
measurement, i.e., $H_C(T=0)=0$.
Our aim will be to construct a self-consistent
procedure to compute the compensation term $H_C$
to any order in $\lambda$: 
\beq
H_{C} = \Delta H^{(0)} +\sum_k \lambda^k \Delta H^{(k)} . 
\eeq
The terms $\Delta H^{(k)}$  are the analogous  
compensations to the back-reaction
up to the  k'th order in $\lambda$. 

To this end, let us expand the solution for $\x(t)$ as
\beq
\x(t,g) = \xzero + \lambda\xone(t) + \lambda^2 \xtwo(t) + \cdots
\eeq
and at each order separate the undisturbed solution,
denoted as $\x^{(k)}_0$, from the non-linear error,
denoted as $\Delta x^{(k)}$:
\beq
\x^{(k)}(g) = \x^{(k)}_0(t) + \Delta \x^{(k)}(t,g), 
\eeq
i.e. for $g=0$ we have $ \Delta \x^{(k)}(t,g=0)=0$ and 
$\x^{(k)}(g=0)= \x^{(k)}_0(t)$
is the solution without the coupling to a measuring device.
 
Likewise, the shift of the output register , $\delta\P = \P(T) - \P(0)$, 
will be expanded in powers of $\lambda$, and at each order 
we shall evaluate the terms of the decomposition:
\beq
\delta\P^{(k)} = \delta \P^{(k)}_0 + \Delta\P^{(k)}.
\label{errors}
\eeq
Here, $\Delta\P^{(k)}$ is the error due to the back reaction to
the  k'th order.
The knowledge of $\Delta\P^{(k)}$ will allow us to construct 
the  appropriate compensation 
$\Delta H^{(k)}$, up to the same order $k$.
 
To proceed we shall now simplify the problem, 
and let $n=3$ in eq. (\ref{model}). 
It can be shown however that the following will be valid for 
any general non-linear potential. 

To evaluate $\Delta\P^{(k)}$ we first use
eqs. (\ref{intrep}) to obtain an integral solution to any 
order:   
\beq
\xzero(t,g) = \x_0(t) + \Q \int^t_{0}g(t')D(t-t') dt', 
\eeq
\beq
\xone(t,g) = \int^t_0 \xzero(t')\xzero(t')D(t-t') dt' ,
\eeq
\beq
\xtwo(t,g) = 2\int_0^t \xzero(t')\xone(t') D(t-t') dt' , 
\eeq
etc., where $\x_0$ is the free solution ($g=\lambda=0$), and 
\beq
D(t) = {\sin\Omega t\over \Omega}.
\eeq

The result for the zeroth order was already obtained above, 
where we 
found 
\beq
\Delta H_0= {1\over2} k \Q^2 ,
\eeq
where $k$ is the c-number given by eq. (\ref{k}).
Here, and in the following we shall ignore any potential 
problems with ordering of non-commuting  operators.

To the first order in $\lambda$ we obtain
\beq
\delta \P^{(1)} = g_0 \int^T_0 dt \int_0^t [\x^{(0)}]^2 dt' .
\eeq
Thus,
$$
\Delta \P^{(1)} = g_0 \int^T_0 dt \int_0^t 
\biggl[2\alpha(t')\Q\x_0(t') 
+ \alpha^2(t')\Q^2 \biggr] dt'
$$
$$
= g_0 \int_0^\infty \biggl[2\alpha(t')\Q\x_0(t') 
+ \alpha^2(t')\Q^2 \biggr] dt' \int_0^T\theta(t-t') dt
$$
$$
= g_0 \int_0^\infty \biggl( (T-t')\theta(T-t')
-(-t')\theta(-t')\biggr) \biggl[2\alpha(t')\Q\x_0(t') 
+ \alpha^2(t')\Q^2 \biggr] dt'
$$
$$
= g_0 \int_0^T (T-t') 
\biggl[2\alpha(t')\Q \Bigl( \x^{(0)}(t') - \Q \alpha(t') \Bigr ) 
+ \alpha^2(t')\Q^2 \biggr]dt'
$$
\beq
=  g_0 \int_0^T (T-t') 
\biggl[2\alpha(t')\Q  \x^{(0)}(t') dt' -
\alpha^2(t)(T)\Q^2\biggr]dt' .
\label{dpone}
\eeq
where $\alpha(t)$ is defined in 
eq. (\ref{alpha}).
As could have been anticipated, the new feature of the 
error terms up to the 
first order in $\lambda$, is their depends on 
the variable  $\x$ of the system.
The last term does not depend on system variables and therefore
may be trivially compensated as a spring-like compensation. 
(Alternatively, as discussed in Section 3., 
we may use a second measuring device to measure the combination 
${1\over g_0T}\P - k\Q - k'(T)\Q^2$, where $k'$ 
can be found by integrating  the last term in eq. (\ref{dpone}).
)

Equation (\ref{dpone}) suggests adding as 
a compensation the term
\beq
\hone =
g^2(t) (T-t) \biggl[\alpha(t)\Q^2\x(t) 
- {1\over3}\alpha^2(t)\Q^3 \biggr] .
\label{hc1}
\eeq

With the compensation $\hzero+\lambda\hone$ added
to the Hamiltonian,
the eqs. of motion  become
\beq
\dot \P = g(t)\x - k(t) \Q  
+\lambda g^2(t) (T-t) \biggl[2\alpha(t)\Q\x(t) 
- \alpha^2(t)\Q^2 \biggr] ,
\label{picor}
\eeq  
\beq
{d^2\x\over dt^2} + \Omega^2 \x =  \lambda \x^2 +
g(t)\Q 
+ \lambda g^2(t) (T-t) \alpha(t)\Q^2  . 
\label{xcor}
\eeq

The new term on the right hand side of (\ref{xcor}) 
modifies  
the solution for $x (t)$ to 
\beq
\x(t) = \x_0(t) + \alpha(t)\Q  + \lambda\beta(t)\Q^2 
+ \lambda\int_0^T \x^2(t')dt' .
\eeq
The first order is modified 
by the $\beta(t)\Q^2$ which can trivially be compensated
by adding also the term ${1\over3}\lambda g(t)\beta(t)
\Q^3$. This completes  the compensation to the first order.
It is  important  to note  that the a first order
compensation 
can not modify lower order compensations but only higher orders.
This allows us to go to higher orders in $\lambda$ without effecting the
corrections
found at lower orders. 

Let us see how one can proceed to higher orders. 
To the second order we obtain:
\beq
{d{\Delta\P}^{(2)}\over dt} = g(t)\Q\Delta \x^{(2)} 
+2g^2(t)(T-t)\alpha(t)\Q \Delta\x^{(1)}(t),
\eeq
where
\beq
\Delta\x^{(1)}(t) = \beta(t) \Q^2  + 
\int^t_0 \Delta [\x^{(0)}(t')]^2dt',
\eeq
and 
\beq
\Delta\x^{(2)}(t) = 2\int^t_0 \Delta[\x^{(0)}(t')\x^{(1)}(t')]
dt'
\eeq
where the notation  $\Delta[\cdots]$,  on the right hand side
in the eqs. above, indicates that only  
$\Q$-dependent terms of $[\cdots]$ are included.

The computation can be carried out, as before. 
The only technical subtlety is that now 
$\Delta \P^{(2)}$ contain also triple integrals over time like:
$\int dt \int x_0(t') dt'\int p_f(t'') dt''$.
To proceed one has  to reduce this integral to 
a single integration over time, 
or eliminate the integrals completely. 
Otherwise, the compensations would have to be non-local in time.
In our case of a first quantized theory,  we can always use 
the property of linear operators (but not field operators):
$\pf(t)$ (or  $\xf(t)$) can always  be expressed
as a linear combination of $\xf(t=0)$ and $\pf(t=0)$.
Therefore, we can integrate over free operators
and reduce:
$\int_0^{t'} \pf(t'')dt'' =
c_1(t')\xf(t')+c_2(t') \pf(t')$. The 
integral above can be hence  reduced to 
a single integral of the form 
$\int (c'_1(t)\xf^2 + c_2'(t)\xf \pf) dt$.
Finally we find that the compensation to the second order 
has the following form:
\beq
\htwo =\Q^2( \gamma_{1}\lbrace \x, \p\rbrace +
\gamma_{2} \x^2) +\Q^3(\gamma_3 \x +
\gamma_4 \p ) + \gamma_5 \Q^4 . 
\eeq
where $\gamma_{i}(T)$ are generally time dependent c-numbers. 
In principle it seems that this  procedure can be carried out 
 to any  order in $\lambda$.

\section{Break-Down of the perturbative approach}

Although by using a perturbative 
approach we can formally 
evaluate  self-consistently 
compensation terms to any order 
in the nonlinear coupling constant, this approach 
still does not allow precise measurements. 
Unlike classical measurements, in the limit of a precise
measurement,
the variable $\Q$, which is conjugate to the ``output register''
$\P$,  becomes completely uncertain. 
Since the expansion for the compensation $H_C$ 
generally depends on $\Q$, for a given $T$, 
the validity of the perturbative
expansion must break-down beyond a certain precision 
$\Delta\x_{min}$.

To show this consider the first order compensation term. 
From eq. (\ref{hc1}) we have 
$\Delta H^{(1)} \approx T\alpha(T) \Q^2 \x$.
A necessary (but not sufficient)
condition for a well behaved expansion is therefore: 
\beq
\lambda T\alpha(T) \Q^2 \x \sim \lambda T^3 \Q^2 \x << 1 .
\eeq
Substituting a rough approximation 
$\Q^2 \sim (\Delta\Q)^2$ and $\x\sim \Delta x$
and using $\Delta \P \sim  \Delta \x/g_0 T$ we obtain 
the condition
\beq
\Delta \x > \Delta \x_{min} \approx  \lambda g^2_0 T^5 .
\eeq
Only by letting $T\to0$ or $\lambda \to 0$ we regain 
$\Delta\x_{min} \to 0$.

\section{Discussion}

In this article we have examined the measurability of 
observables by means of a continuous measurement for  
linear and non-linear theories. 
We have seen that as long as the exact non-perturbative solution 
is known it is possible,  at least in principle,  to device an 
exact measurement. The formal answer to the two questions asked 
in the introduction is therefore  positive. 
 
Nevertheless, the surprise is that without the knowledge of 
the exact non-perturbative 
solution, a perturbative approach can not be used to obtain 
arbitrary accurate measurements. 
For any non-vanishing value 
of the non-linear coupling constant, and a given interaction
time $T$, the perturbation breaks-down for any measurement
with accuracy better than 
$\Delta\bar\A < \Delta\bar\A_{min}(T,\lambda)$.
The origin of this failure is that 
the perturbation to any order, necessarily involves 
terms which depend on $\Q$,
the conjugate to the output variable $\P$. 
Therefore for sufficiently precise measurements, $\Delta \P$, 
can  be arbitrarily small, and $\Delta \Q$ becomes arbitrarily large. 
At a certain accuracy the perturbation  is bound to break-down.

This conclusions clearly applies to the question 
of measurability of an observable at an instant by means of a continuous 
measurement. If the exact solution is not known, 
we shall not be able to compensate perturbatively for the error.
Thus, under the restriction of finite time measurements, 
a preparation  of a certain {\it pure state} 
$\psi$ of a non-linear system,  requires the knowledge of 
the exact non-perturbative solution of the  equations of motion.

\vspace {2 cm} 

{\bf Acknowledgment}

B.R.  would like to thank W. G. Unruh for valuable
discussions, and to A. Peres and L. Vaidman for helpful remarks.

\end{document}